\documentclass{mn2e}
\usepackage{epsfig}

\newif\ifAMStwofonts
\AMStwofontstrue

\title[Microlensing signatures of starspots]
{The microlensing signatures of photospheric starspots}

\author[M.~A. Hendry, H.~M. Bryce and D. Valls--Gabaud]
{Martin A. Hendry$^{1}$, Helen M. Bryce$^{1,2}$ and 
David Valls--Gabaud$^{3}$\\
$^1$ Dept. of Physics \& Astronomy, University of Glasgow, Glasgow G12 8QQ, 
UK\\
$^2$ Dept. of Physics and Astronomy, 203 Van Allen Hall,
University of Iowa, Iowa City, IA, 52242, USA\\
$^3$ CNRS UMR 5572, Observatoire Midi-Pyr\'en\'ees, 14 Avenue E. Belin, 
31400 Toulouse, France}

\date{Accepted 2002 March 12.
      Received 2002 March 11;
      in original form 2001 June 18}

\pagerange{\pageref{firstpage}--\pageref{lastpage}}
\pubyear{2002}

\begin{document}
\maketitle

\label{firstpage}

\begin{abstract}

Point lens microlensing events with impact parameter close to the
source stellar radius allow the observer to study the surface
brightness profile of the lensed source.  We have examined the effects
of photospheric star spots on multicolour microlensing lightcurves
and investigated the detectability of such spots in different
wavebands as a function of spot temperature, position, radius and 
lens trajectories.  We include the effects of limb darkening and spot
projection as a function of position on the stellar disk. In
particular we apply the updated, state-of-the-art `NextGen' 
stellar atmosphere models
of Hauschildt \emph{et al.} which predict very strong limb darkening,
and which are likely to be applicable to the source stars considered here.

Our results indicate that star spots generally give a clear signature only 
for transit events. Moreover, this signature is strongly suppressed by limb 
darkening for spots close to the limb, although the spots may still be clearly
detected for favourable lens trajectories. 

It is also clear that intensive temporal sampling thoughout the
duration of the transit is necessary in order for such events to be
effective as a tool for imaging stellar photospheres.  Nonetheless,
with sufficiently well sampled light curves of good photometric
precision, microlensing can indeed place useful constraints on the
presence or otherwise of photospheric starspots.

\end{abstract}


\section{Introduction}
Recently gravitational microlensing has been developed as a
tool to probe the
distribution and nature of dark matter in the Milky Way (see
Paczy\'{n}ski 1996 \nocite{Pac96} 
for a detailed review), with several hundred candidate events detected to date.

In the past few years, several authors have
established that analysis of microlensing events can, in principle,
provide much useful
information about
the star being lensed, in the particular case where the star has
significant angular extent. 
Such `finite source effects'  were considered initially for a uniform
circular disk (see e.g. Bontz 1979\nocite{Bontz79}; Nemiroff 
and Wickramasinghe 1994\nocite{NemWic94}; Gould 1994,
1995\nocite{Gould94}\nocite{Gould952}; Witt and Mao
1994\nocite{WittMao94}; Peng 1997 \nocite{Peng97}). These calculations
demonstrated that the
finite extent of the source would produce a deviation from the
standard lightcurve for a point source, allowing estimation of the
source radius and hence the Einstein radius of the lens. 
Heyrovsk\'{y} and Loeb
(1997)\nocite{HeyLoe97} extended this treatment to the microlensing of
a uniform elliptical source. 

The microlensing signature of a non-uniform disk (due to, for example,
limb darkening) was considered by
Valls-Gabaud (1994)\nocite{VG94}, Witt (1995)\nocite{Witt95},
Bogdanov and Cherepashchuk (1995, 1996) and Simmons, Newsam and Willis
(1995)\nocite{Simmons95b} and
Simmons, Willis and Newsam (1995)\nocite{Simmons95a}, 
who showed that the microlensed
lightcurves would display a chromatic signature as the lens
effectively sees  a star of different radius at different
wavelengths.  Further studies of the chromatic signature of
extended sources followed in, e.g. Gould and Welch (1996)\nocite{GouWel96},
Valls-Gabaud (1996)\nocite{VG96}, Sasselov (1997) \nocite{Sas97}
and Valls-Gabaud (1998)\nocite{VG98}. These authors discussed the 
possibility of using microlensing to constrain stellar atmosphere
models by determining the radial surface brightness profile of the
lensed star as a function of wavelength. Hendry et
al. (1998)\nocite{Hendry98} investigated a non-parametric approach to
inverting the radial surface brightness profile from multi-colour
lightcurves, using the Backus-Gilbert method (see also Gray and Coleman
2002\nocite{Gray00}).  

The calculations of Simmons, Newsam and Willis (1995) were carried out 
for a grey model atmosphere with a linear limb darkening law;  this
simple model was improved in Valls-Gabaud (1998) to incorporate  
linear, quadratic and logarithmic limb darkening laws, with
coefficients calculated for the Johnson wavebands $U$ to $K$.  More
recently Heyrovsk\'{y}, Sasselov and Loeb (2000) \nocite{HeySasLoe99}
have used one particular model atmosphere for a red giant star to explore
the efficacy of high-precision photometric and spectroscopic
microlensing data as a detailed probe of red giant atmosphere models.

The study of stellar atmospheres is one of several specific
astrophysical applications of extended 
source events which have been discussed in recent literature. Others
include the measurement of stellar rotation (e.g. Maoz and Gould
1994\nocite{MaozGou94}), diagnosis of stellar winds (Ignace and Hendry
1999\nocite{Ignace99}), and detection of extra-solar planets (Gaudi and
Gould 1997\nocite{GauGou97a}).

A subject which has received comparatively little attention to date
is the microlensing signatures of \emph{non-radial} surface brightness 
profiles on the source disk - as would result from e.g. the presence of
photospheric starspots. Sasselov (1997) presented light curves for a
simple model of a cool circular spot on the photosphere of a red giant
star.  This treatment was extended in Heyrovsk\'{y} and Sasselov
(2000), hereafter HS00, 
to consider the signatures of hot and cool circular spots as a
function of spot position, lens impact parameter, stellar radius and
spot area. The authors presented maps of ``spot detectability'',
adopting a change in the microlensed flux of 2\% 
as the criterion for detectable features in that particular model
atmosphere of a red giant. Han et al. (2000) extended
this analysis for the case of caustic crossings produced by a binary lens, 
but did not consider the effects of limb darkening nor used model
atmospheres.

The stellar atmosphere model adopted in HS00 -- based on those developed in
Heyrovsk\'{y}, Sasselov and Loeb (2000) \nocite{HeySasLoe99} -- 
included the effects of limb darkening, which -- as one would expect -- 
suppresses the signature of a hot or cool feature close
to the stellar limb.  The numerical procedure adopted both by HS00 
and Han et al. (2000), 
however, ignored 
geometrical foreshortening which would result in a circular spot
appearing 
as an
ellipse in projection, as the spot is displaced from the centre of the
disk. 

In Bryce, Hendry and Valls-Gabaud (2002)
we improved the geometrical treatment of HS00 and Han et al. (2000) 
by introducing an
accurate model of the effects of foreshortening for a
circular spot plus an exploration of the effects of limb darkening 
within the spots
themselves as opposed to the uniform spots in these two papers.  Here 
we extend the analysis of that paper to include a more comprehensive
exploration of the effects of non-radial surface brightness variations 
as a function of source, spot and lens parameters. In particular we extend our
model atmospheres calculations to incorporate the recent `Next
 Generation' (NextGen) 
atmosphere models of Hauschildt et al. (1999a,b), which include a more
comprehensive treatment of spherical geometry and the effects of molecular 
opacity in the outer atmospheres of cool giants.

The \emph{gravitational imaging} of stellar photospheres is a potentially very
useful tool for stellar astrophysics.  The atmospheres of
cool giants are relatively poorly understood -- not least because of
their highly evolved and thus theoretically complex
nature. Fundamental issues such as their typical rotation speeds
remain uncertain, and in fact spectroscopic observations of
microlensing events have already been proposed as a useful diagnostic
of red giant rotation (Gould 1997\nocite{Gould97}).

Detecting photospheric spots in ways different from, e.g., rotationally
induced photometric modulation or Doppler imaging 
could bring considerable insight on their properties.
For instance it is not clear what the covering factor could be,
particularly if many spots are present which bias the determination
of unspotted surface properties. In Doppler imaging analyses the temperature 
of the spots is usually cooler than the photosphere of the stars under study, 
typically by 600 -- 1200 K,
but this may be at least in part due to selection effects. 
There is an interplay
between spot area and temperature which microlensing could perhaps
remove. Although Doppler imaging has revealed irregularly-shaped
spots, there are still many uncertainties in the precise
interpretation of such maps.

The precise role of, e.g., convection in the outer atmospheres of red
giants is also unclear.  Semi-regular variables, such as $\alpha$
Orionis, undergo brightness changes of about one-half magnitude on a
timescale of years; this variability has been associated with the
intermittent appearance of large convection cells on the photosphere.
HST observations of  $\alpha$ Orionis do indeed reveal evidence of a
single unresolved bright area on the photosphere (Gilliland and Dupree 
1996)\nocite{GilDup96} although subsequent spectroscopic observations
suggest that its origin may be due not to convection but rather to
``an outwardly propagating shock wave'' (Uitenbroek, Dupree and
Gilliland 1998)\nocite{Uitenb98}. 

The study of starspots is clearly, therefore, an important aspect of
gravitational stellar imaging.  Moreover, microlensing provides a
unique opportunity for probing such features on the surface of (most
probably) slowly rotating red giants, which are completely unsuitable
candidates for the powerful Doppler Imaging technique (see e.g.
Strassmeier and Linsky, 1996, and references therein). 

Finally, we also note that when stars are resolved by a lens, the 
probability of planet detection may be reduced since the effect of the
finite source is to smear out, and thus suppress, the planetary signature
(see for instance Vermaak, 2000, for a recent analysis). Photospheric
starspots could mimic such features, and hence provide an unexpected
noise background for planet detection via microlensing, although the time
of observation of the features and their chromatic signature should generally
provide a suitable means of discriminating between spots and planets.

The structure of this paper is as follows.  In \S2 we discuss the Next
Generation stellar atmosphere models used in our calculations, in \S3
we discuss our model for the 
microlensing of an extended source and in \S4 we provide some informative
examples of the lightcurves produced by spotted stars and explore the
``detectability'' of starspots as a function of e.g. size, position
and temperature.  In \S5 we
present our conclusions.
Appendix A describes in detail our geometrical treatment of a circular spot.

\section{NextGen model atmospheres}

Computations of stellar surface brightness profiles have been carried out for
several decades but until very recently were based on an approximate treatment
whereby the Planck function was used to compute central intensities, $I_0$, for
different wavebands, and the intensity, $I(\mu)$, as a function of
(the cosine of)
emergent angle, $\mu$, was then given by a simple parametric model,
such as the linear model, namely: 
\begin{equation}
I(\mu) = I_0 \left [ 1 - c_1 (1-\mu) \right ]
\end{equation}
The coefficient $c_1$  depends on the temperature, gravity
and chemical composition of the source and on the wavelength of
observation. Although several improvements have been made in recent
years (see Valls-Gabaud, 1998, for further references) it is
clear that some of the underlying hypotheses are not realistic.

The recent `Next Generation' stellar atmosphere models,
computed by
Hauschildt and collaborators (1999a,b), considerably improve upon these
simplistic models
in several important respects. The calculations are carried out
assuming spherical
geometry for giant stars, and the intensity
profiles are computed directly --
without assuming a Planck law or a parametric model for the dependence
on emergent
angle. Moreover, the intensity calculations are based on a huge
library of atomic and molecular 
lines, with about $2 \times 10^8$ molecular lines contributing to a
typical giant atmosphere 
model at $T_{\rm eff} = 3000$ K.

\begin{figure*}
\centerline{\psfig{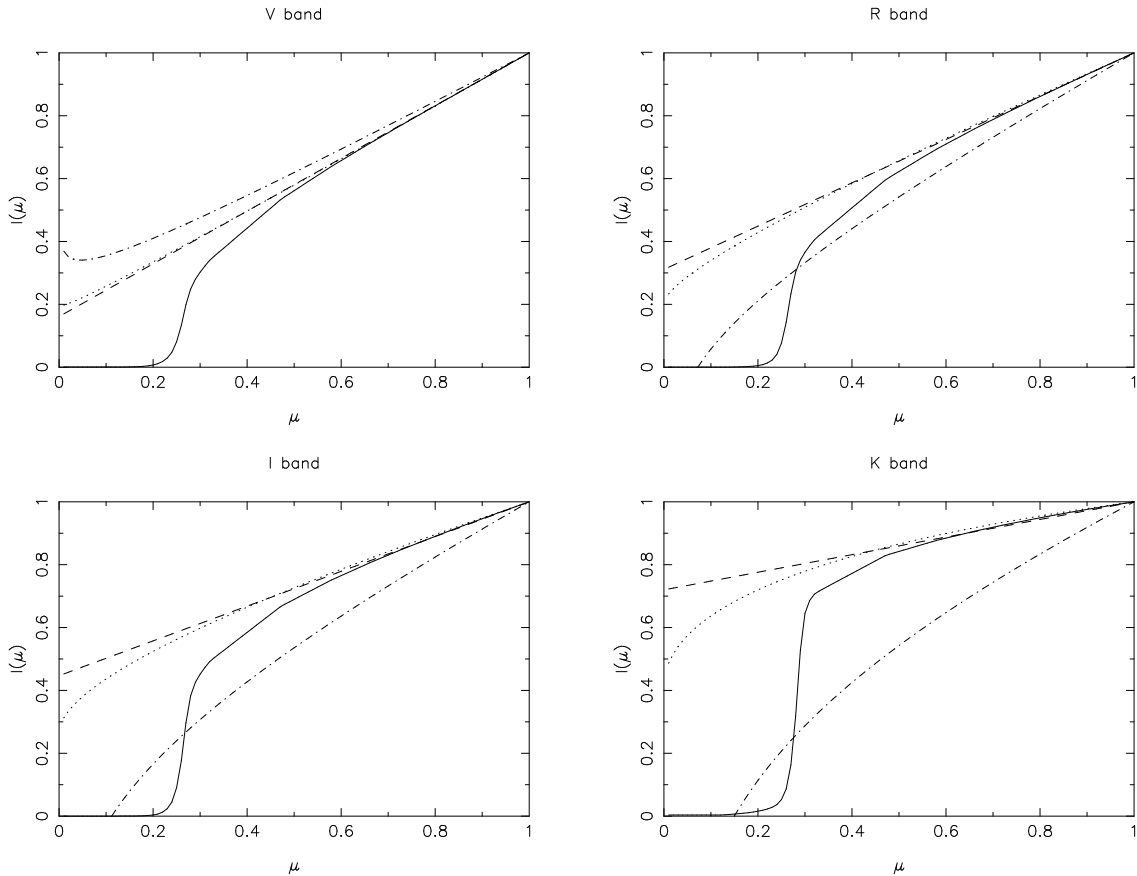}}
\caption[]{Intensity profiles for a giant star of 
$T_{\rm eff} = 4250$K and $\log g = 0.5$, in four Johnson colour bands: 
$V$, $R$, $I$ and $K$. The solid curve shows the NextGen profiles, 
while the dashed, dash-dotted and dotted curves denote the linear, 
logarithmic and square root models respectively.}
\label{ng1}
\end{figure*}

The dramatic difference in the dependence of limb darkening on emergent 
angle between the traditional models and NextGen models is illustrated in 
Figure \ref{ng1}. This figure shows the intensity profiles for a giant star of 
$T_{\rm eff} = 4250$K and $\log g = 0.5$, in four Johnson colour bands: $V$, 
$R$, $I$ and $K$. The solid curve shows the NextGen profiles, while the 
dashed, dash-dotted and dotted curves denote the linear, logarithmic and 
square root models respectively. 
We can see from this figure that there is a sudden decrease in the intensity 
of the NextGen models as one approaches the limb of the star --
i.e. at $\mu \simeq 0.2$. This feature
arises from the adoption of spherical geometry and the 
improved modelling of molecular scattering in the
outer atmosphere of the star, and is 
clearly an effect which one would expect to be highly relevant to
stellar atmospheres probed by 
microlensing, but is completely absent from the earlier
parametric models (e.g. Kurucz, 1994) which predict significant intensity
all the way to $\mu =0$. 

This strong limb feature was detected during the recent EROS
2000--BUL--5 event, where a NextGen model was able to fit the 
$H\alpha$ equivalent width variation across the photosphere of a K
giant (Albrow et al. 2001b), improving the earlier
calculations by Valls--Gabaud (1996, 1998). Accordingly the results 
presented in this paper 
are calculated using NextGen atmosphere models rather than parameterised limb
darkening laws.

\section{Microlensing of an extended spotted source}

The majority of candidate
microlensing events are well fitted by a point source, point lens
model. The characteristic scale for microlensing is usually taken to
be the Einstein radius of the lens, defined as

\begin{equation}
R_{E} = \sqrt{\frac{4GM}{c^{2}}\frac{(D_{S}-D_{L})D_{L}}{D_{S}}}
\label{eq:einsteinradius}
\end{equation}

Here $M$, $D_{S}$, $D_{L}$ are the lens mass, the distance to source
and the distance to lens respectively.
The angular Einstein radius, $\theta_{E}$, is

\begin{equation}
\theta_{E}=\frac{R_{E}}{D_{L}}=\sqrt{\frac{4GM}{c^{2}}\frac{(D_{S}-D_{L})}
{D_{L}D_{S}}}
\label{eq:aer}
\end{equation}

In the case of a point source being lensed by a point mass the
amplification, $A$, of the source depends only on the projected angular
separation, $u$ of the lens and source.

\begin{equation}
A(u)=\frac{u^{2}+2}{u\sqrt{u^{2}+4}}
\label{eq:pointamplif}
\end{equation}
and hence is independent of wavelength.
 
However when the angular stellar radius is comparable to the angular
Einstein radius of the lens (i.e. an extended source) the integrated,
lensed flux from the stellar disk is then given by

\begin{equation}
F(u) = \int_{s=0}^{R} \int_{\theta=0}^{2 \pi} \,
I(s,\theta) A(u) \, s \, ds d\theta
\label{eq:lensedflux}
\end{equation}

where the source angular radius, $R$, and angular separation, $u$, are
measured in units of $\theta_{E}$.  The amplification factor, $A(u)$, is
simply the point source, point lens amplification given by
Equation~\ref{eq:pointamplif}. 

For the background, unspotted photosphere we assume that
$I_{*}(s,\theta)=I_{*}(s)$ -- i.e. the surface brightness profile is
radially symmetric, and these values are then taken from the NextGen
model atmospheres, for the appropriate waveband under examination and
the  effective temperature and $\log g$ of the source being modelled.

Our procedure for integrating the flux difference due to starspots is very
similar to the above, except that the appropriate temperature at which we 
evaluate the surface brightness now varies as a function of the polar angle, 
$\theta$. Thus
\begin{equation}
\Delta F(u) =
\int_{\cal{A}} \, \left[ I_{\rm{sp}} (s, \theta) - I_{*} (s) \right]
\, A(u) \, s \, ds d\theta
\label{eq:deltaf}
\end{equation}
where $\cal{A}$ is the projected area of the disk covered by spots. We may 
apply Eq.~\ref{eq:deltaf} for the case of several starspots by simply
adding together the contribution, $\Delta F(u)_i$, from each spot --
provided only that none of the
spots overlaps with any other. Furthermore we also apply the NextGen
limb darkening to the spots modelled, i.e. spots close to the limb
will not be as bright as spots of the same effective temperature in
the centre of the photosphere.  
The key problem is, then, a purely geometrical one: 
to determine the $(s,\theta)$ coordinates marking the boundary of the area 
covered by each starspot at 
any given time. A similar problem was dealt with by Dorren (1987),
but we describe our method  fully in Appendix~A.
The only physical simplification we have made is to assume that
the effective gravity within each spot is the same as the
gravity in the photosphere. (It would be computationally
straightforward -- but prohibitively time-consuming -- to 
allow for variations in effective gravity within each spot).
Although there is as yet no direct
observational evidence to support this assumption (Donati,
private communication) we can show that it is reasonable by the
following physical argument.  Assuming that the growth of spots
is regulated by horizontal pressure gradients created by the
evaporation of magnetic flux tubes, Gray (1992) quotes 
the approximate relation, $g_{\rm spot}^{0.6} P_{\rm spot} = 
\rm{constant}$, between the spot effective gravity and pressure -- 
i.e. the larger the spot pressure, the lower its effective gravity.
Relating $P_{\rm spot}$ to the stellar magnetic field and
inserting typical values for a solar-type star with
$\log g = 4.5$, one finds $\log g_{\rm spot} \sim 3.6$.  For
giant stars, with a smaller magnetic field, the difference
in $\log g$ should be smaller than about 0.2 dex, and so
the small effect of varying effective gravity within the
spot can be neglected.

\section{Results}
\subsection{Illustrative examples}
\begin{figure*}
\centerline{\epsfig{file=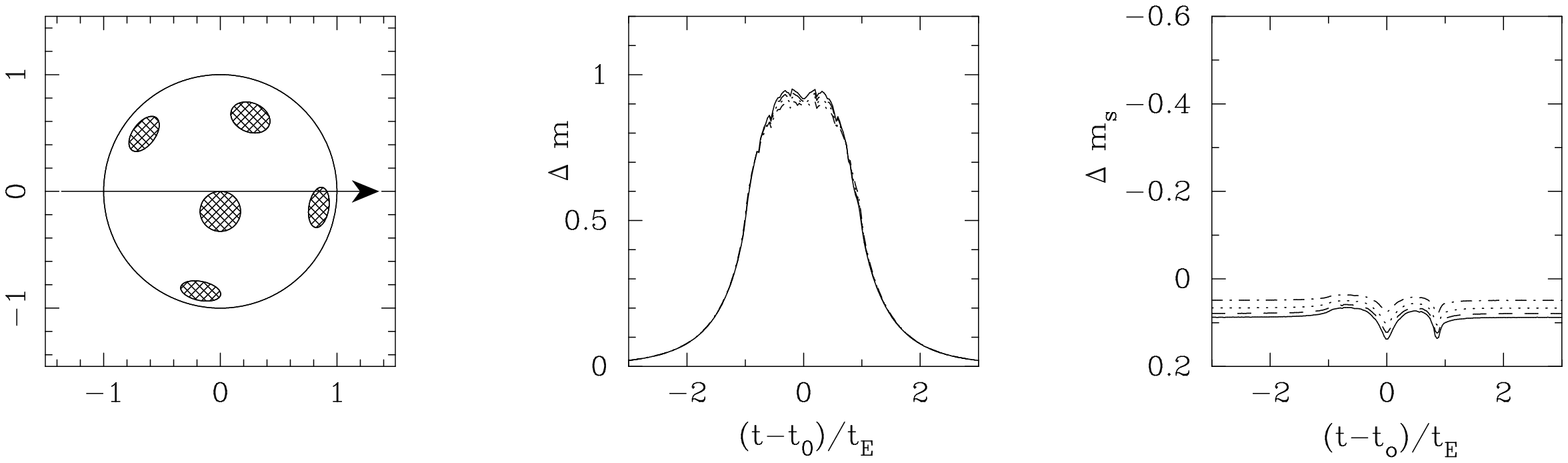,
angle=-90, width=16cm}}   
\caption{\small{The $V$, $R$, $I$ and $K$ microlensing light curves
produced by the transit of a lens, with impact parameter
$u_{0} = 0.0 \theta_{E}$, across the disk of a $5000$ K  star, of radius
1 $\theta_{E}$, with $\log g = 4.0$, with five $4200$ K starspots (as shown in
the left hand panel)
of radius $10^{\circ}$.  The magnitude changes, $\Delta m$ and $\Delta
m_{\mathrm{S}}$ are as defined in Equations \ref{deltam} and \ref{deltams}.}}
\label{cool1}
\end{figure*}
\begin{figure*}
\centerline{\epsfig{file=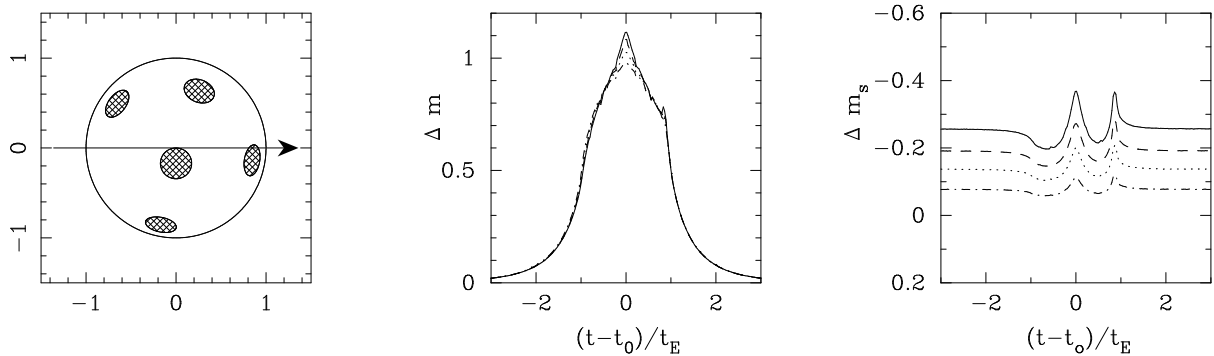, angle=-90,
width=16cm}}
\caption{\small{The $V$, $R$, $I$ and $K$ microlensing light curves
produced and represented in the same manner as Figure
\ref{cool1} but  with five hot $4800$ K starspots on the disk of a
$4000$ K, $\log g = 1.0$ star.  }}
\label{hot1}
\end{figure*}
\begin{figure*}
\centerline{\epsfig{file=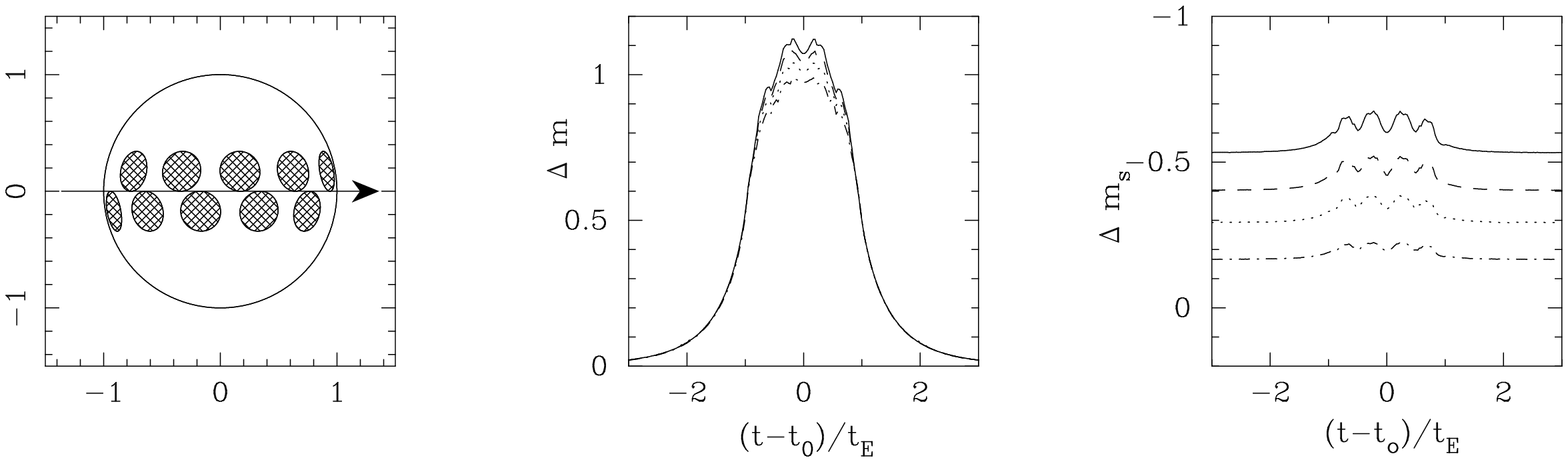,
angle=-90, width=16cm}}
\caption{\small{The $V$, $R$, $I$ and $K$ microlensing light curves
produced by the transit of a lens, with impact parameter
$u_{0} = 0.0 AER$, across the disk of a $4000$ K star, of radius
1 $\theta_E$, with $\log g = 1.0$,
with a band of $4800K$ starspots (as shown in the left hand panel)
of radius $10^{\circ}$ viewed at an inclination of $90^{\circ}$.  The
magnitude changes, $\Delta m$ and $\Delta
m_{s}$ are as defined in Equations \ref{deltam} and \ref{deltams}. }}
\label{band1}
\end{figure*}
\begin{figure*}
\centerline{\epsfig{file=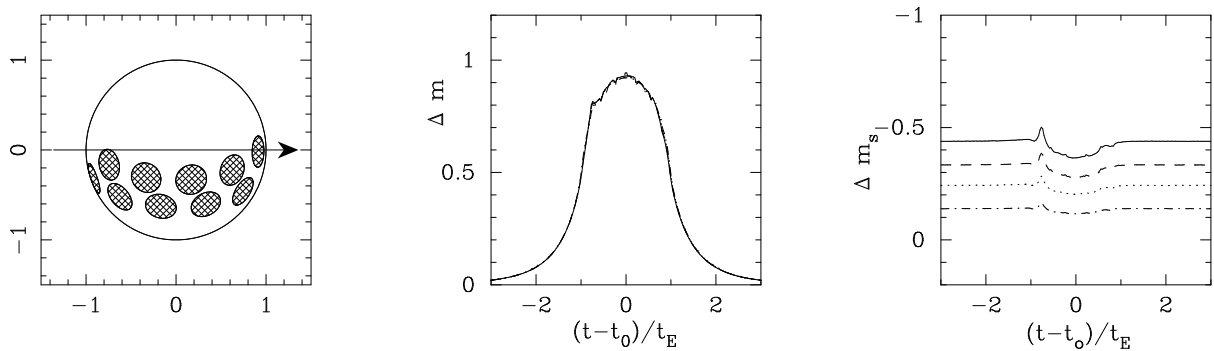,
angle=-90, width=16cm}}
\caption{\small{The $V$, $R$, $I$ and $K$ microlensing light curves
of the same source as in Figure \ref{band1} but now viewed at an
inclination of $60^{\circ}$. }}
\label{band2}
\end{figure*}

Figures \ref{cool1} to \ref{band2} present illustrative lightcurves for 
several
different configurations of star, lens and spots. In each of these
figures we have assumed that the angular radius of the source is equal
to the angular Einstein radius of the lens. While this choice 
of stellar radius is observationally unrealistic and produces low
amplification events, for illustrative purposes it has the advantage
of producing spot features on the lightcurves which are clearly
distinguishable by eye. We will consider somewhat more realistic values of 
the ratio of stellar radius to Einstein radius later in this section.
Lightcurves were computed from the NextGen models in the $U$ to $K$ bands; 
however for clarity only 4 bands are shown in the following figures. The
$V$, $R$, $I$ and $K$ bands are represented by a continuous, a dashed, 
a dotted and dashed-dotted line respectively.
 
In Figures \ref{cool1} to \ref{band2} three panels are shown. The left 
hand panel shows
the stellar disk, illustrating the position and size of the starspots
and the trajectory (indicated by an arrow) of the lens during the
event (in fact all four cases are for a lens impact parameter,
$u_{0} = 0$).  The middle panel indicates the 
(absolute value of the) change, $\Delta m$,
in apparent magnitude as a function of time, as given by the formula

\begin{equation}
\Delta m = \left| 
2.5 \log_{10}\left(\frac{F_{\mathrm{LSP}}}{F_{\mathrm{USP}}}\right) \right|
\label{deltam}
\end{equation}
where $F_{\mathrm{LSP}}$ and $F_{\mathrm{USP}}$ denote the flux from the
lensed, spotted
star and unlensed spotted star respectively. (Note that, since
the effect of lensing is -- of course -- to amplify the unlensed flux,
the sign of the change in apparent magnitude is alwys negative).
Similarly the right hand
panel shows the change, $\Delta m_{\mathrm{S}}$ in apparent magnitude 
resulting specifically from the presence of the spot(s), as given 
by the formula

\begin{equation}
\Delta m_{\mathrm{S}} = - 2.5 \log_{10}\left(\frac{F_{\mathrm{LSP}}}
{F_{\mathrm{LSF}}}\right)
\label{deltams}
\end{equation}
where $F_{\mathrm{LSF}}$ denotes the lensed flux from the spot-free star.
(Note that $\Delta m_{\mathrm{S}}$ may be either positive or negative).

While the middle panel indicates the time evolution in observed
apparent magnitude of the stellar disk plus spot(s) during the
microlensing event, it is difficult to isolate from this panel the
contribution to the lightcurve of the spots themselves.  Thus,
although the final panel indicates a flux ratio which would not be
directly observable, it nevertheless makes clear the residual
magnitude change between the observed flux from a spotted source
and the \emph{predicted} flux from an event with identical stellar and
lens parameters (i.e. stellar radius, atmospheric limb darkening,
lens impact parameter etc) but \emph{without} the non-radial surface
brightness variations due to the presence of spots.

Note that $\Delta m_{\mathrm{S}}$ as defined by Equation \ref{deltams}
would give a
wavelength dependent non-zero offset, even in the absence of lensing.
This arises because the integrated unlensed flux from the spotted star
would, in any case, introduce a magnitude change when compared to the
unspotted star. Thus we can think of 
$\Delta m_{\mathrm{S}}$ as composed of two
contributions;
\begin{equation}
\Delta m_{\tiny{S}} = \Delta m_{\tiny{SL}} - 2.5 \log_{10}
\left(\frac{F_{\tiny{USP}}}{F_{\tiny{USF}}}\right)
\end{equation}
where $F_{\mathrm{USF}}$ denotes the unlensed flux from the spot-free
star. Hence $\Delta m_{\mathrm{SL}}$ can be regarded as the magnitude change
due to the \emph{lensing} of the spots, while the second term
represents the non-zero offset discussed above.  Although we are, of
course, primarily interested in $\Delta m_{\mathrm{SL}}$, it is useful to
include the non-zero offset in the right hand panels of Figures 
\ref{cool1} to \ref{band2}
as it adds to the clarity of the lightcurve deviations.

Figure \ref{cool1} illustrates the $V$, $R$, $I$ and $K$ lightcurves produced
by the transit of a lens across the disk of a $5000$ K dwarf star and five
$4200$ K circular spots of radius $10^{\circ}$. 
These spot parameters are motivated by, and are broadly consistent
with, the results of stellar maps derived from Doppler imaging 
studies of active dwarfs (see e.g. Strassmeier and Linsky, 1996, 
and references therein). It can be seen that
the two cool spots very close to the lens trajectory are the only
features which contribute significantly to the lightcurve.  The central
spot produces a noticeable `dip' in the  central portion of the
lightcurve, although the feature becomes less prominent at longer
wavelengths and is barely detectable in the $K$ band.  This is not
surprising as the contrast in limb darkening between $5000$ K and
$4200$ K diminishes at longer wavelengths.  The spot close to the right
hand limb of the star produces a signal of slightly smaller amplitude
-- although with a similar dependence on wavelength -- which is due in
part to the geometrical foreshortening clearly visible in the left
hand panel.

Note also that the position of this spot, in the `wings' of the light
curve, renders it very difficult to isolate in the middle panel,
although its presence is obvious in the right hand panel.

Figure \ref{hot1} shows the same lens trajectory and configuration 
of five spots
as in Figure \ref{cool1}, but now with a temperature of $4800$ K on a $4000$ K 
giant star. Here the spot radius and temperature difference is motivated
by the HST observations of $\alpha$ Orionis (Gilliland and
Dupree 1996). Once again, only the two spots which lie very close to the lens
trajectory are `imaged' by the lens, producing a peak in the
lightcurves, which diminishes in amplitude at longer wavelengths. Note
from the right hand panel that the unlensed offset between different
wavebands and the peak magnitude change due to lensing are both
somewhat larger than in Figure \ref{cool1}; i.e. the unlensed (and lensed)
contrast of hotspots is greater than that of cool spots.  This is not
surprising, since what is important is the (log) ratio of spotted to unspotted
{\em flux\/}; this translates to a larger magnitude change for the hot 
spot than for the cool spot, although the temperature difference has the 
same absolute value in both examples.

The lensing signature of multiple spots is also interesting,
particularly if spots are restricted to a given latitudinal
band. Figures \ref{band1} and \ref{band2} show a $4000$ K giant star with a 
large group of
$4800$ K,  $10^{\circ}$ radius spots filling a $40^{\circ}$ band in
latitude, straddling the equator.  In Figure \ref{band1} the inclination of 
the source star is $90^{\circ}$, so that the lens trajectory intersects the
spot bands.  We can
see from the middle and right hand panels that the spots produce a
complex pattern, with a series of `bumps' each one of which is roughly
identifiable with an individual spot.  The amplitude of
each feature, however, varies with spot longitude -- due to
foreshortening and limb darkening -- and also diminishes considerably with
increasing wavelength as before.  

Another interesting feature is
apparent in the right hand panel of Figure \ref{band1}.  We can see in 
the wings of the light curve that even in the absence of lensing the 
presence of the large band of spots produces, as one might expect, a 
substantial magnitude offset which is also strongly wavelength dependent 
(more than 0.5 mag. at $V$, reducing to less than 0.2 mag. at $K$). 
In the central part of the lightcurve, as the lens transits the star and
crosses the band of spots, the flux in each waveband is amplified due to
lensing. Moreover the effect of the lensing is also to increase very 
slightly the {\em differential\/} magnitude offset in the central part of 
the light
curve, compared with the offset in the wings. However, this differential
effect is slightly smaller {\em between\/} the series of 'bumps'
(when the lens is passing through a 'gap' in the band of spots -- e.g. at
$t = t_0$) than it is at the peaks of the bumps (when the lens is at its
closest to one or more of the spots in the band).  Thus, the increase in flux
due to the presence of the spots (which is augmented by the lens but which
would be apparent even if there were no lensing) is slightly diluted when the
lens is amplifying the flux from the background star more than the spots 
themselves.

This interplay between the amplified flux from the spots and the background 
star is more clearly seen in the corresponding panels of Figure \ref{band2}, 
which shows the microlensing light curves of the same band of spots, but
now on a star of inclination $60^{\circ}$. Again we see a wavelength-dependent
magnitude offset even when there is no lensing, due to the presence of the
hot spots. Shortly after the stellar transit begins there is a small 'spike' 
in the light curves, as the lens passes very close to one of the spots.
Thereafter, however, there is a large dip around the central part
of each light curve, which also slightly reduces the differential
magnitude offset between the wavebands.  This is because, during this time, 
the flux is dominated by the amplified contribution from the region of the 
star very close to the lens, which -- as can be seen from the left hand
panel of Figure \ref{band2} -- is spot free.  Thus we see that even when the
lens does \emph{not} pass directly over a spotted region, the presence
of spots elsewhere on the star still results in a clearly detectable
microlensing signature in the lightcurve as a whole.

The systematic deviations from the unspotted light curves illustrated
in Figures \ref{cool1} to \ref{band2} highlight a potential pitfall of 
estimating stellar
and lens parameters from discretely sampled data. In Figure \ref{hot1}, for
example, the peak produced by the central spot feature could -- if
sampled with poor time resolution -- lead to the estimation of a
stellar radius which was systematically smaller than the true radius.
In this respect, in addition to the obvious benefit of increasing the
time resolution of observations, multiwavength photometry can also be
effective in breaking the apparent degeneracy between a spotted star
and a spot-free star of smaller radius.  This is because the effect of
the starspot is considerably reduced at longer wavelengths.  Similar
remarks apply to the case of the cool spots on Figure \ref{cool1}, although
since the amplitude at a given wavelength of the spot feature is
somewhat smaller for cool spots, the (positive) systematic bias in the
estimated stellar radius would be less significant than the (negative)
bias obtained from higher amplification hotspots.

\subsection{Investigating spot detectability}

\begin{figure*}
\centerline{\epsfig{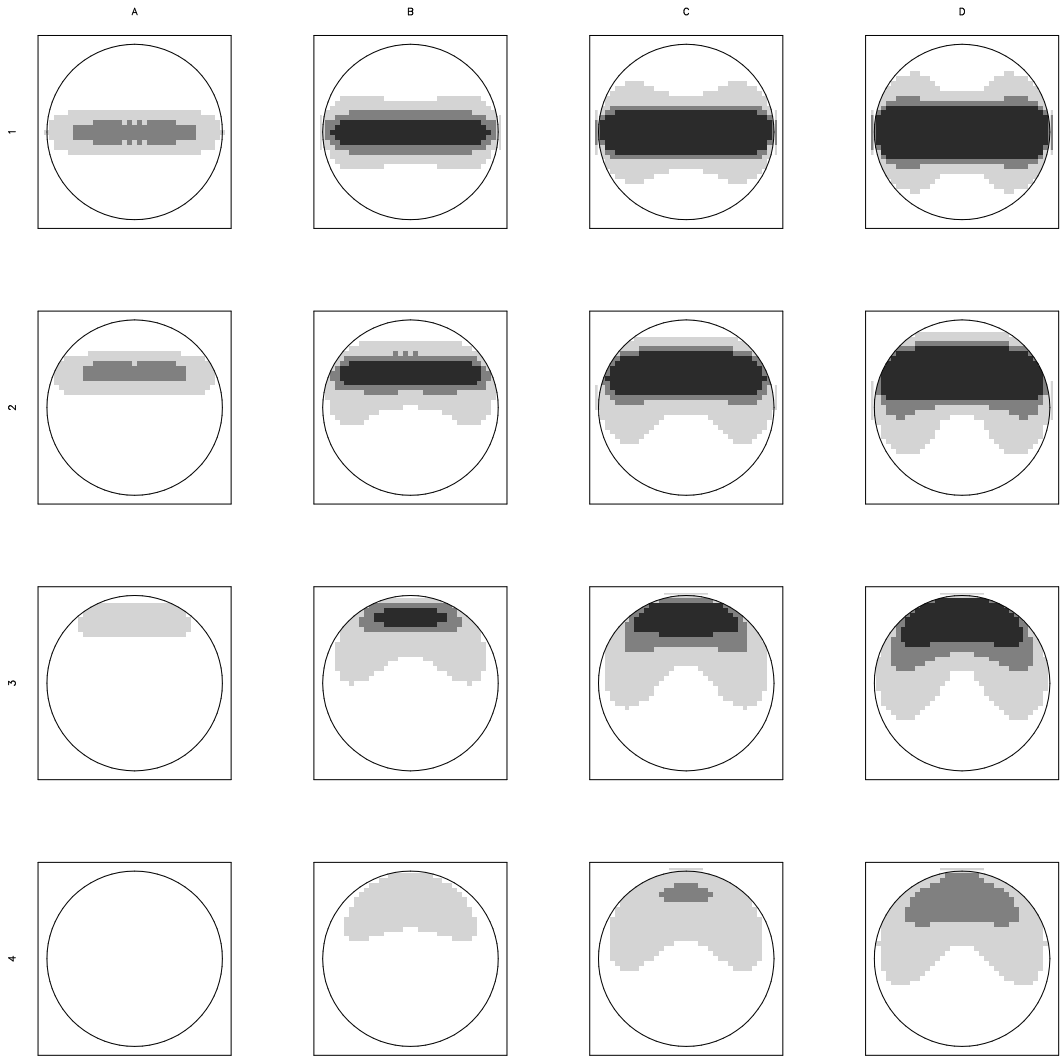}}
\caption{\small{The regions of excess magnification of $0.01, 0.05$
and $0.1$ magnitudes from the lightest gray to darkest due to a spot
feature of $4800$ K on a 0.1 $\theta_E$, $\log g = 1.0$, $4000$ K star.  Rows
from top to bottom
correspond to lens trajectories
with minimum impact parameter, $u_{0} = 0.0, 0.4, 0.8$ and 
$1.2$, respectively, normalised to the source radius. Columns from
left to right correspond to spots of radius $3^{\circ}, 6^{\circ},
9^{\circ}$ and $12^{\circ}$ respectively.}}
\label{con1}
\end{figure*}

We next carried out an extensive investigation of spot detectability
as a function of spot radius and position and lens impact parameter.
These calculations were performed for the somewhat more 
realistic case of a star with angular radius $R_* = 0.1 \theta_E$. 
This value is comparable to -- albeit slightly larger than --
the estimated stellar radius for MACHO 95-30 (Alcock et al. 1997),
and is similar to the value adopted in the calculations of HS00.
There could, of course, be some merit in considering sources of
smaller angular radius since MACHO 95-30 represents the largest
source star detected to date. Moreover, a smaller source star would
also yield a larger microlensing signature. However, given the
reduced probability of a point lens transit for such a star --
together with the greatly reduced likelihood of intensive 
photometric monitoring during that transit -- we adopt
$R_* = 0.1 \theta_E$ as a reasonable compromise between 
considering a star with unrealistically large angular radius and 
a star with unrealistically low probability of being observed 
undergoing a point lens transit. (See also Section 5 below).

Figure \ref{con1} presents $V$ band contour maps of spot detectability for a
range of spot and lens parameters.  In all cases, the background
stellar temperature and spot temperature were taken to be $4000$ K and
$4800$ K respectively.  Rows from top to bottom correspond to lens trajectories
with impact parameter, $u_{0} = 0.0, 0.4, 0.8$ and 
$1.2$, respectively, normalised to the source radius. Columns from
left to right correspond to spots of radius $3^{\circ}, 6^{\circ},
9^{\circ}$ and $12^{\circ}$ respectively.

We assume for each light curve that a total of 80 equally spaced
observations were made during $t_E$, i.e. the time taken for the lens
to move 1 angular Einstein radius.  (Hence, for example $t_E =
40$ days would imply a sampling rate of two observations per day).

The contours shown denote the contribution to the change in magnitude
arising specifically from the lensing of the spot, defined as $\Delta
m_{\mathrm{SL}}$ in Equation 9 above.  The darkest level shown denotes a peak
magnitude change of at least 0.1 mag.(i.e. at least one observation
for which $\Delta m_{\mathrm{SL}} \ge 0.1$); the lightest level denotes a
peak change of 0.01 mag.

It is clear from these contour maps that, as one might expect, the
detectability of starspots is improved greatly in regions where the spot
lies close to the trajectory of the lens.

Nevertheless, at least for the larger spots, a substantial fraction of the
stellar surface would give a detectable signature if a starspot were
present at that location.  For a spot of radius $12^\circ$ and a lens impact
parameter $u_0 = 0$, for example, we can see from the top right panel of
Figure 6 that the spot would provide a magnitude change of 0.1 mag. or 
greater over about 30\% of the visible disk of the star.  We can compare
that with the area covered by the spot itself as a fraction of the area of 
the visible hemisphere, that is $\sin^2 \delta$, so in this case about
4.3\%. Hence microlensing is very efficient at detecting spots under these
conditions.

It is interesting to note, however, that even for the case of
$u_{0} = 0.12 \theta_{E}$ (i.e. impact parameter slightly greater than
the stellar radius) one still finds that approximately $40$ \% of the
stellar disk yields a magnitude change in excess of 0.01
mag (compared with the 4\% area covered by each spot). 
Although such a signal would clearly be difficult to detect, its residuals
are comparable to the accuracy levels achieved by the best current
photometry in microlensing follow up surveys.

The furthest left column of Figure \ref{con1} (spot radius of $3^{\circ}$)
shows, as expected, the smallest `detectable' regions of the
photosphere.  Indeed, for $u_{0} = 0.12 \theta_{E}$ it can be seen that
no detectable signal would be obtained.  Note also the discrete
features in the top left panel (and in a few other panels) which arise
because of the small size of the spots and the discrete nature of the
the temporal sampling.  With more intensive time coverage these
features would become more smooth.

It should be stressed, of course, that the contour maps shown in Figure 6
assume that one can predict accurately the {\em unspotted\/} lightcurve --
i.e. one can accurately estimate the source radius and lens impact
parameter from the lightcurves as a whole.  Uncertainty in those parameters
would generally impair the sensitivity of the lightcurves to the presence
of spots by reducing the detectability area of the contour maps.  A detailed
investigation of the impact of `global' parameter errors on the estimation
of linear limb darkening coefficients from analysis of event OGLE-99-BUL-23
was carried out by Albrow et al. (2001a), and a similar statistical analysis
would be important in the context of starspots 
when confronting real microlensing data.  Global parameter errors would
clearly have the most damaging effect for small spots, and indeed might
`wash out' completely the signature of $3^\circ$ spots in the leftmost
column of Figure 6.  In those regions where the spot signature exceeds
0.1 mag., however, spots would still be easily detectable by current
folow up surveys, even allowing for uncertainties in the global
parameters.

\subsection{The feasibility of gravitational imaging}

\begin{figure*}
\centerline{\epsfig{file=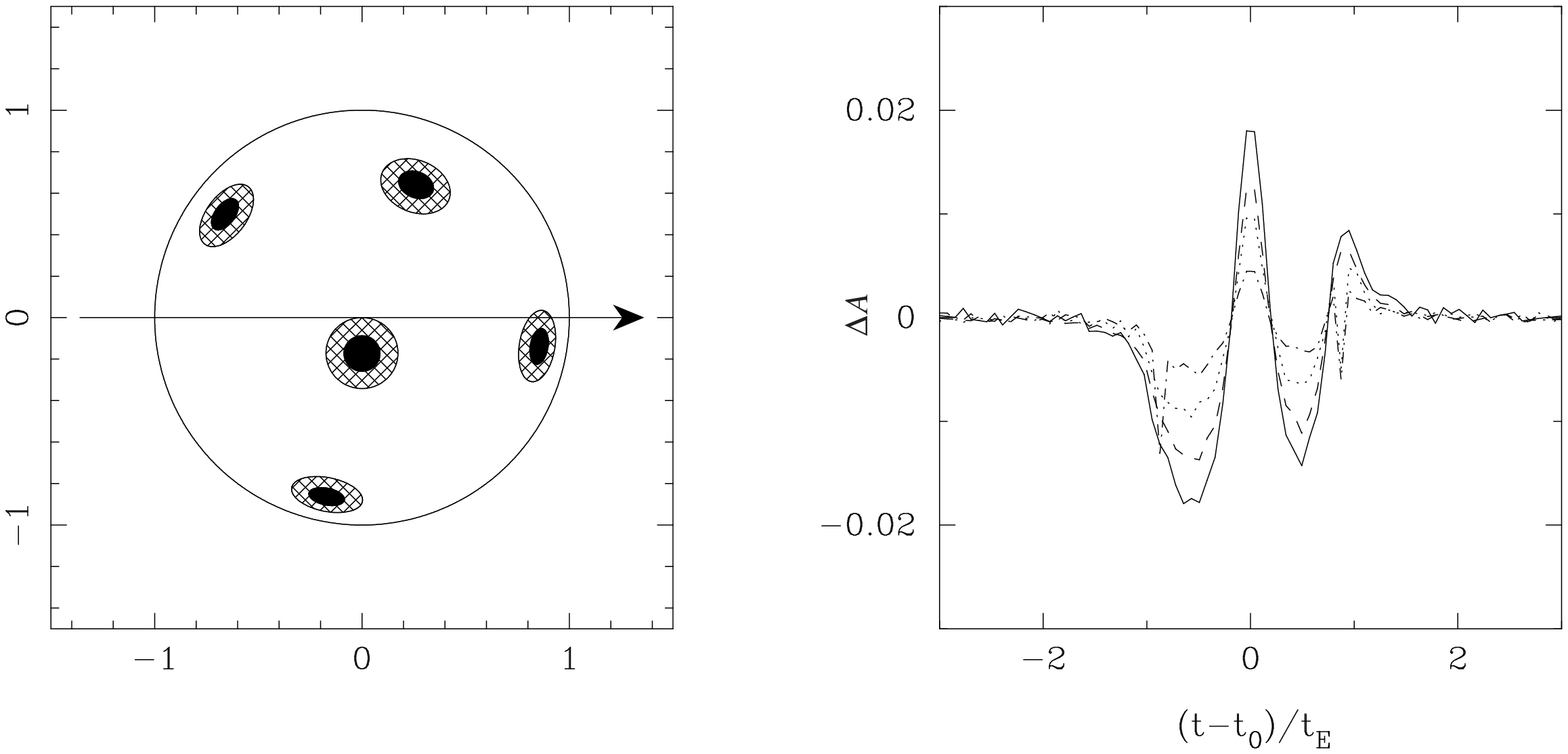,angle=-90,width=14cm}}
\caption{\small{A comparison of the $V$, $R$, $I$ and $K$ microlensing
light curves
produced by the transit of a lens, with impact parameter
$u_{0} = 0.0 AER$, across the disk of a $4000$ K star, of radius
1 $AER$, with $\log g = 1.0$,
with five $4800$ K starspots
of radius $10^{\circ}$ (i.e. Figure 2) against an identical source
other than additional central temperature structure
at $5600$ K of $5^{\circ}$ (as shown in the left hand panel) within
the spots. $\Delta A$ is as defined in Equation 10.}}
\label{struc1}
\end{figure*}

We have seen in the previous section that high time resolution
observations during a transit event can place useful constraints on
the existence or otherwise of spots on particular regions of the
photosphere.  The question remains as to whether one can constrain the
detailed structure of spot features from such observations.

Figure \ref{struc1} shows a comparison between the light curves produced by the
event illustrated in Figure \ref{hot1} and a similar event, with the same
stellar and lens parameters, but in which the spots have additional
temperature structure -- specifically a central `umbra' of temperature
$5600$ K and radius $5^\circ$ surrounded by a cooler `penumbra' of
temperature $4800$ K (the same temperature as the uniform spots of
Figure \ref{hot1}).  The right hand panel of Figure \ref{struc1} shows 
the difference, in
the amplification between the two scenarios, given by
\begin{equation}
\Delta A = \left( \frac{F_{\mathrm{LSP}}}{F_{\mathrm{USP}}}
\right)_{\rm struc}
\;  - \; 
\left(\frac{F_{\mathrm{LSP}}}{F_{\mathrm{USP}}}\right)_{\rm no-struc}
\end{equation}
It is evident from Figure \ref{struc1} that the effect of the spot structure is
small, with a peak deviation from the uniform temperature case of only
$\Delta A \simeq 0.02$, which corresponds to $\Delta m \simeq 0.01$
mag. Clearly, then the detection of temperature structure, given the
accuracy of current photometry, would be very difficult.  A more
serious difficulty, however, is presented by the severe ill-posedness
of the problem: since spots need not in general be circular it is
likely that the specific photometric signature of temperature
structure within a circular spot could be closely approximated by a
non-circular spot of uniform temperature. In essence, one cannot
effectively constrain the 2 dimensional structure of a given spot feature from
only a 1 dimensional microlensing light curve.  Similar remarks
clearly apply to the photosphere as a whole, where groups of
(arbitarily shaped) individual spots could mimic the signature of a
single, larger, spot and vice versa.

Possible rotation of the source star (which is certainly an area of
some interest in the study of red giants and supergiants) is another
potential source of degeneracy in the signatures of starspots.
Depending on the orientation of the rotation axis with respect to the
lens trajectory, a small rotation of the source during the transit of
a spot of radius, say, $10^{\circ}$ might broaden or narrow the spot
signature in the light curve by an amount equivalent to a change in
the spot radius of about $1^{\circ}$.  Clearly the magnitude of the
effect depends on the star's rotation period and the lens timescale,
but the main point is that -- without a precise model for the shape
and size of the spot feature -- one cannot accurately constrain the
rotation period from the spot signature.  It is interesting to
consider whether the use of multicolour lightcurves might at least
partially break the degeneracy between models of spot structure and/or
stellar rotation.  The practical limitation here, however, is the fact
that at longer wavelengths spot signatures are considerably supressed
and are less sensitive to surface temperature and other stellar
parameters.  In future work, we will, nevertheless, investigate
optimal methods for combining multicolour data to constrain atmosphere
and spot models in this manner.

\section{Discussion}
The results presented in Section 4 clearly show that the impact of
non-uniform surface brightness profiles on the microlensing light
curves of point caustic events is, in principle, detectable with
current observational precision.  There are several important caveats,
however.

Firstly, as discussed in Section 4.3, the degenerate nature of the
microlensing light curve, which is a convolution over the stellar
disk, renders the task of constraining the precise number, position,
shape, size and temperature contrast of starspot(s) an ill-posed
problem.  Notwithstanding these limitations, however, Figure 6 shows
that -- for a transit or near transit event -- a significant fraction
of the photosphere will yield an observable magnitude change due to
the presence of even a single spot, provided it has sufficient area
and temperature contrast. (In fact, since the important quantity in
determining the size of the spot signature is the ratio of spotted to
unspotted flux, changing the spot temperature contrast is a more important
factor than changing the spot area -- at least in the wavebands considered
here; for the idealised case of black-body radiation and ignoring limb
darkening, the spot flux will be proportional to the product of the area
and the fourth power of the temperature).
 
Thus it is clear that the \emph{failure} to detect a significant
deviation from the light curve signature expected for an unspotted
source does indeed allow one to place robust limits on the fraction of
the stellar disk (i.e. the `filling factor') which could be covered by
starspots.

Of course the crucial issue here is not
simply the presence or otherwise of spot features, but also one's
ability to detect them.

As we discussed in Section 4.2, 
an important consideration here is the question of how well determined
are the `global' parameters of the event -- i.e. the lens minimum impact
parameter and the source radius which determine the predicted 
{\em spot-free\/} lightcurve. These can, in principle, 
still be determined from the overall light curve shape, and indeed the use of 
multicolour observations -- which we have shown can be an important 
diagnostic of the properties of the spots -- can also improve the accuracy 
with which the global parameters are determined. However, any 
significant uncertainty in the global parameters would weaken 
the constraints on the spot features since the predicted spot-free light curve 
would be less well determined. A rigorous statistical treatment of the impact
of errors in the global parameters would be a crucial step before robust
limits on the properties of spots could be inferred from real microlensing
data.

The main practical limitation, however, to the use of point lens events
as probes of stellar surface features is simply the 
trajectory of the lens. Spots which
lie in regions of the disk far from the lens path will not be readily
detected under any circumstance.  Even regions lying very close to the
lens trajectory,
however, would require excellent photometry and high temporal
sampling, in order to detect smaller features.  In the illustrative
example of Figure \ref{con1} -- for which we assumed a sampling rate of
e.g. approximately twice per day for an event with timescale $t_E =
40$ days -- spots smaller than $3^{\circ}$ in radius were
undetectable even with a minimum detection criterion of having only
one observation with $\Delta m_{\mathrm{s}} \ge 0.1$ mag.   Clearly a
statistically convincing detection might require several
consecutive data points with $\Delta m_{\mathrm{s}}$ greater than
some chosen threshold.  This could be achieved either through having
larger spots (as is indeed the case for the other columns of Figure 6)
or by increasing the temporal sampling.  While the latter is, of
course, always desirable, telescope logistics dictate that -- without
prior warning that a transit is about to occur -- a sampling rate
greater than about twice per day is not realistic.  Transits of a
point mass lens are rare and even for high amplification
events, difficult to predict from the pre-transit lightcurve. A more
`observationally friendly' scenario is the transit of an extended
source by the caustic structure produced by a binary lens. This is
because in such a high amplification microlensing event it becomes
necessary to treat every source as an extended source and hence the
source's surface brightness profile can be investigated. Furthermore
the intial transit `into' the structure acts
as a alert, allowing intensive observations of the second crossing
`out of' the structure to be planned. Such `alert response' photometry
is currently being carried out with the aim of detecting low-mass
companions but clearly has scope to image stellar atmospheres, allowing
the detection or otherwise of photospheric starspots. Gravitational 
imaging by non-point lens objects will be addressed in a forthcoming paper.

\section*{Acknowledgments}

The authors wish to thank John Simmons and Rico Ignace for many useful
discussions, and the referee, Penny Sackett, for a number of helpful
suggestions.  HMB acknowledges a PPARC studentship.

\appendix

\section{Starspot geometry and integration limits}

\subsection{Coordinate systems}

We calculate the integrated flux from the star, in both the lensed and 
unlensed case, in terms of an integral over the projected stellar disk. There
are three coordinate systems relevant to this calculation:

\begin{enumerate}
\item{
$(\alpha' , \phi')$ : spherical polar coordinates on the surface of the star,
with the stellar equator defining $\phi' = 0$, and with $\alpha'$ measured
counter-clockwise from the direction which is co-planar with the star's
rotation axis and the line of sight (as shown in Fig. A1).
}
\item{
$(\alpha , \phi)$ : spherical polar coordinates on the surface of the star, but
with polar axis $(\phi = {\pi \over 2})$ defined as the projection of the
star's rotation axis on the plane of the sky and with azimuthal angle, 
$\alpha$, measured counter-clockwise from the line of sight (as shown
in Fig. A2).
}
\item{
$(s , \theta)$ : projected circular polar coordinates on the stellar disk
(i.e. the plane of the sky), with $\theta$ measured counter-clockwise from
the $y$-axis (see below).
}
\end{enumerate}

Figures A1 and A2 illustrate these coordinate
systems and their associated Cartesian coordinate axes. Thus, we
define the $x$-axis to be the line of sight, the $z$-axis to be the
projection of the star's rotation axis onto
the plane of the sky, and the $y$-axis to be the direction which completes
a right-handed coordinate system. It is then easy to see that the $y$-axis
and $y'$-axis are identical, and the $x'$-axis and $z'$-axis are obtained
from the $x$-axis and $z$-axis by a rotation of $( {\pi \over 2} - i )$ about
the $y$-axis, where $i$ is the inclination of the star.
In summary, for a star of radius, $R$, and a general point $(x,y,z)$ on the
stellar surface
\begin{eqnarray}
x & = & R \cos \alpha \cos \phi \, \, = \, \,
R ( \cos \alpha' \cos \phi' \sin i + \sin \phi' \cos i ) \nonumber \\
y & = & R \sin \alpha \cos \phi \, \, = \, \,
R \sin \alpha' \cos \phi' \nonumber \\
z & = & R \sin \phi \, \, = \, \,
R ( \sin \phi' \sin i - \cos \alpha' \cos \phi' \cos i )
\label{A1}
\end{eqnarray}
where all coordinates are expressed in units of the angular Einstein radius
(AER) of the lens.
\begin{figure}
%\vspace{7cm}
 \begin{center}{
   \epsfxsize 5.0 true cm
    \leavevmode
   \epsffile{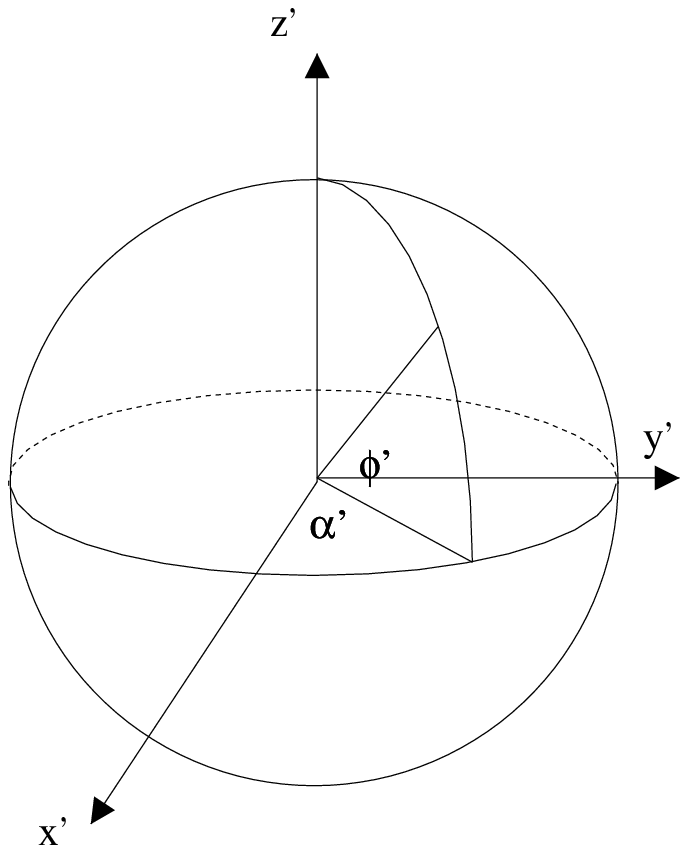}
  }\end{center}
\label{figA1}
\caption{Spherical polar coordinates on the surface of the star,
with the stellar equator defining $\phi' = 0$, and with $\alpha'$ measured
counter-clockwise from the direction which is co-planar with the star's
rotation axis and the line of sight.}
\end{figure}
\begin{figure}
%\vspace{7cm}
 \begin{center}{
   \epsfxsize 5.0 true cm
    \leavevmode
   \epsffile{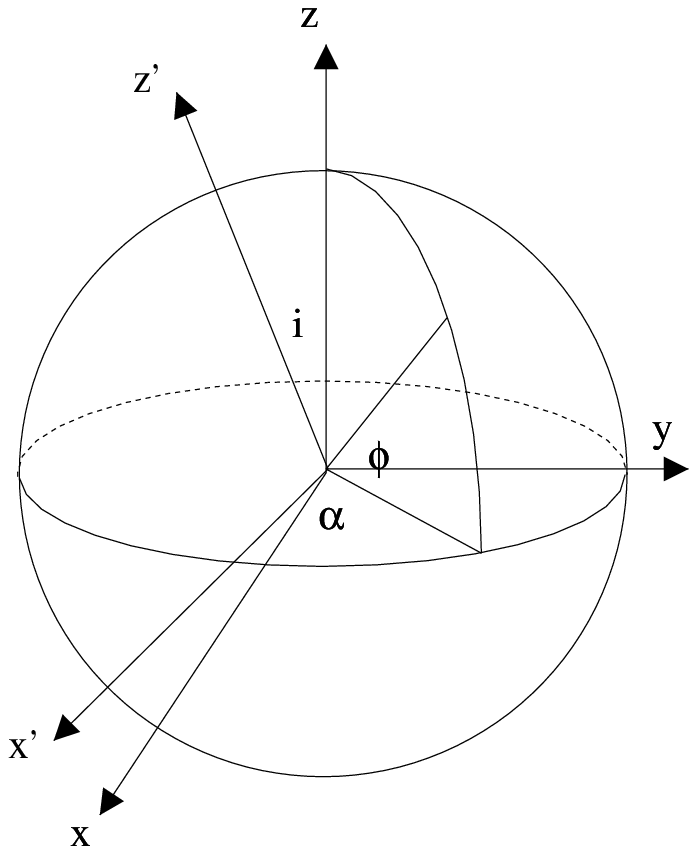}
  }\end{center}
\label{figA2}
\caption{Spherical polar coordinates on the surface of the star, but
now with polar axis $(\phi = {\pi \over 2})$ defined as the projection of 
the star's rotation axis on the plane of the sky and with azimuthal angle, 
$\alpha$, measured counter-clockwise from the line of sight}
\end{figure}
Consider now the path of a point lens, as seen in projection on the sky. 
Figure \ref{figA3} shows the lens trajectory and the position of the
lens at some general point, $L$, and at time, $t$. Here, $u$
denotes the impact parameter of the lens and $u_{0}$ denotes the
impact
parameter at the time of closest approach, $t_{0}$, when the lens has
position angle $\theta_{0}$, as shown. Let $t_E$ denote the time for
the lens to move $1\theta_E$. The coordinates 
$(y_{\tiny{\rm{L}}}, z_{\tiny{\rm{L}}})$, of the lens at position,
$L$, are
\begin{eqnarray}
y_{\tiny{\rm{L}}} & = &
u_{0} \cos \theta_{0} - \frac{( t - t_{0} )}{t_E}
\sin \theta_{0} \nonumber \\
z_{\tiny{\rm{L}}} & = &
u_{0} \sin \theta_{0} + \frac{( t - t_0 )}{t_E}
\cos \theta_{0}
\label{A2}
\end{eqnarray}
The projected separation, $u$, of the lens from an arbitrary point, 
$(y,z) = (s \cos \theta , s \sin \theta)$, on the stellar disk is
\begin{equation}
u = \left [ 
( y_{\tiny{\rm{L}}} - s \cos \theta )^2 + 
( z_{\tiny{\rm{L}}} - s \sin \theta )^2 \right ]^{1/2}
\label{A3}
\end{equation}
\begin{figure}
%\vspace{7cm}
 \begin{center}{
   \epsfxsize 5 cm
    \leavevmode
   \epsffile{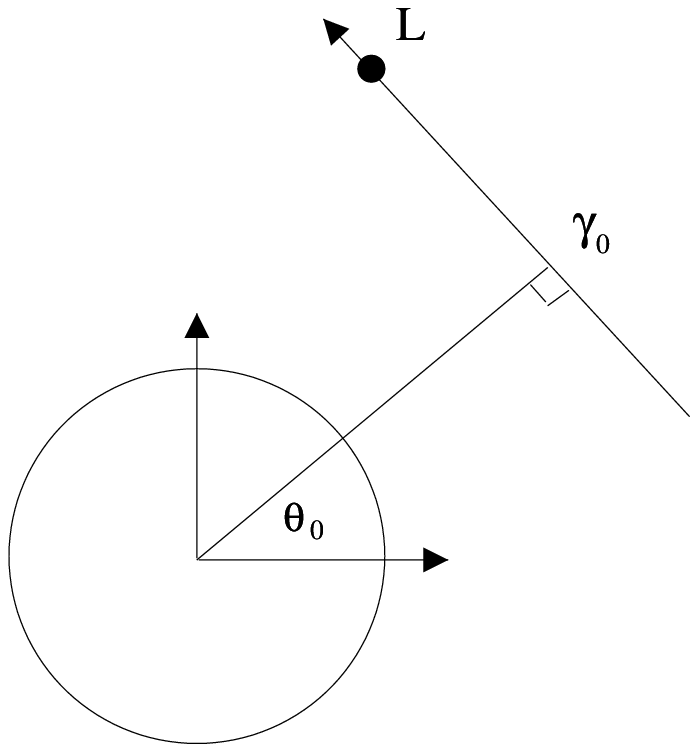}
  }\end{center}
\caption{Lens trajectory}
\label{figA3}
\end{figure}

\subsection{Defining the boundary of a circular starspot}

We consider circular starspots -- i.e. the locus of points defining the
boundary of a spot is a small circle of angular radius $\delta$, say. We
suppose that each starspot maintains constant radius, temperature and
latitude (in the stellar-based coordinate system) throughout the microlensing
event, but that its longitude changes if the star is rotating.

Let $({\alpha_p}' , {\phi_p}')$ denote the (stellar-based) coordinates of the
centre of the starspot. If the star is not rotating, these coordinates remain
fixed; if the star is rotating with period, $P$, and the spot centre transits
at time, $t_0$, then
\begin{equation}
{\alpha_p}' = \frac{2 \pi}{P} ( t - t_0 )
\label{A4}
\end{equation}
We can easily obtain from eqs. \ref{A1} the projected circular polar 
coordinates, $(s_p, \theta_p)$, of $P$ in the observer-based coordinate system.
The circumference of the spot describes a planar circle of radius 
$R \sin \delta$. The centre, $C$, of this circle lies inside 
the star, i.e.
\begin{eqnarray}
x_c & = & R \cos \delta \cos \alpha_c \cos \phi_c \nonumber \\
y_c & = & R \cos \delta \sin \alpha_c \cos \phi_c \,\, = \,\, 
s_c \cos \theta_c \nonumber \\
z_c & = & R \cos \delta \sin \phi_c \,\, = \,\, 
s_c \sin \theta_c
\label{A5}
\end{eqnarray}

Note that $s_c = s_p \cos \delta$ and $\theta_c = \theta_p$.

%\begin{figure}
%\vspace{7cm}
% \begin{center}{
%   \epsfxsize 5.0 true cm
%    \leavevmode
%   \epsffile{figa4.eps}
%  }\end{center}
%\caption{Spot projection}
%\end{figure}

Thus, when seen in projection the centre, $C$, of the planar circle defining
the spot boundary is not coincident with the centre, $P$, of the spot on the
surface of the star, but does lie along the same radial vector joining $P$
to the centre of the stellar disk.

Consider a general point $(x,y,z)$ on the circumference of the spot. We have
\begin{equation}
(x - x_c)^2 + (y - y_c)^2 + (z - z_c)^2 = R^2 \sin^2 \delta
\label{A6}
\end{equation}
and
\begin{equation}
x^2 + y^2 + z^2 = R^2
\label{A7}
\end{equation}
Combining eqs. \ref{A6} and \ref{A7} gives, after some manipulation
\begin{equation}
x x_c + y y_c + z z_c = R^2 \cos^2 \delta
\label{A8}
\end{equation}
which, as expected, defines a plane perpendicular to the position vector
$(x_c, y_c, z_c)$.

\subsection{Spot visibility conditions}

Consider a unit vector in the direction of the spot centre, $C$. i.e.
\begin{equation}
{\hat{n}}_c = (\cos \alpha_c \cos \phi_c, \sin \alpha_c \cos \phi_c,
\sin \phi_c)
\label{A9}
\end{equation}
Let $\eta$ be the angle between the line of sight and ${\hat{n}}_c$. Then
$\cos \eta = \cos \alpha_c \cos \phi_c$. A spot will be fully visible 
provided $\eta \leq {\pi \over 2} - \delta$, i.e.
\begin{equation}
\cos \alpha_c \cos \phi_c \geq \sin \delta
\label{A10}
\end{equation}
Similarly the spot will be fully invisible provided
\begin{equation}
\cos \alpha_c \cos \phi_c \leq - \sin \delta
\label{A11}
\end{equation}
and partially visible when
\begin{equation}
- \sin \delta \leq \cos \alpha_c \cos \phi_c \leq \sin \delta
\label{A12}
\end{equation}

\subsection{Spot centred on the limb of the star}

Suppose first that $x_p = 0$. It is straightforward to show that the spot
circumference appears in projection as a straight line perpendicular to the
radius vector to $(y_c,z_c)$ and the integration limits are
\begin{equation}
\theta_c - \delta \leq \theta \leq \theta_c + \delta \quad \quad \quad
\frac{R \cos \delta}{\cos (\theta - \theta_c)} \leq s \leq R
\label{A13}
\end{equation}

\subsection{Fully visible spot}

Suppose now that $x_c \neq 0$. For a fully visible spot, at any time the 
projected spot will appear as an ellipse centred on $(y_c,z_c)$.
The semi-major axis is perpendicular to the radius vector to $(y_c,z_c)$ 
and some 
straightforward algebra shows that it has length $l_1 = R \sin \delta$.
To determine the semi-minor axis we require to solve for the value(s) of $s$
at which the spot projection intersects the radius vector through $(y_c,z_c)$.
Clearly, at the points of intersection we have
\begin{equation}
y = s \cos \theta_c \quad \quad \quad z = s \sin \theta_c
\label{A14}
\end{equation}
From eq. \ref{A8} it follows that
\begin{equation}
x = \frac{R^2 \cos^2 \delta - y y_c - z z_c}{x_c}
\label{A15}
\end{equation}
Combining eqs. \ref{A7} and \ref{A15} gives
\begin{equation}
(R^2 \cos^2 \delta - y y_c - z z_c )^2 + y^2 x_c^2 + z^2 x_c^2 =
R^2 x_c^2
\label{A16}
\end{equation}
which, substituting from eqs. \ref{A5} and \ref{A14}, may be reduced to the 
quadratic equation in $s$
\begin{equation}
R^2 \cos^2 \delta - 2 s_c s + s^2 - R^2 \cos^2 \alpha_c \cos^2 \phi_c
= 0
\label{A17}
\end{equation}
This has determinant, $\Delta$, which some algebra reduces to
\begin{equation}
\Delta = 4 R^2 \cos^2 \alpha_c \cos^2 \phi_c \sin \delta
\label{A18}
\end{equation}
Hence eq. \ref{A17} has roots
\begin{equation}
s = s_c \pm R \cos \alpha_c \cos \phi_c \sin \delta
\label{A19}
\end{equation}
from which we see immediately that the projected spot ellipse has semi-minor 
axis $l_2 = R \cos \alpha_c \cos \phi_c \sin \delta$.

We can parametrise a general point inside this ellipse as
\begin{equation}
y_E = \omega \, l_1 \, \cos \theta_E \quad \quad \quad
z_E = \omega \, l_2 \, \sin \theta_E
\label{A20}
\end{equation}
Where $0 \leq \omega \leq 1$ and $0 \leq \theta_E \leq 2 \pi$. The coordinates
$(y_E,z_E)$ are related to $(y,z)$ via
\begin{eqnarray}
y & = & y_E \sin \theta_c + z_E \cos \theta_c + s_c \cos \theta_c
\nonumber \\
z & = & z_E \sin \theta_c - y_E \cos \theta_c + s_c \sin \theta_c
\label{A21}
\end{eqnarray}

%\begin{figure}
%\vspace{7cm}
% \begin{center}{
%   \epsfxsize 5.0 true cm
%    \leavevmode
%   \epsffile{figa5.eps}
%  }\end{center}
%\caption{Coordinate system for fully visible spot}
%\end{figure}

The integral in eq. \ref{A7} may then be expressed in terms of $\omega$ and
$\theta_E$, viz
\begin{equation}
\Delta F = l_1 l_2
\int_{\theta_E=0}^{2 \pi} \int_{\omega=0}^{1}
\hspace{-2mm} \left [ I_{\rm{sp}} (s, \theta) - I_{*} (s) \right ]
\, A(d) \, \omega \, d\omega d\theta_E
\label{A22}
\end{equation}

\subsection{Partially visible spot}

The case where a spot is only partially visible is slightly more complicated.
Consider the intersection of the
projected spot ellipse with a circle of radius $s$ on the stellar disk and
centered on $O$. Putting $y = s \cos \theta$, $z = s \sin \theta$,
$y_c = s_c \cos \theta_c$, $z_c = s_c \sin \theta_c$ and substituting in
eq. \ref{A8}, gives, after some further reduction
\begin{equation}
\cos ( \theta - \theta_c ) = \frac{ R^2 \cos^2 \delta - x_c 
\sqrt{R^2 - s^2} }{s s_c}
\label{A23}
\end{equation}
or, writing in terms of $\alpha_c$ and $\phi_c$,
\begin{equation}
\theta = \theta_c \pm \cos^{-1} \left [
\frac{R \cos \delta - \cos \alpha_c \cos \phi_c
\sqrt{R^2 - s^2}}
{s ( \sin^2 \alpha_c \cos^2 \phi_c + \sin^2 \phi_c )^{1/2}} \right ]
\label{A24}
\end{equation}

Thus, for a partially visible spot we integrate eq. \ref{A7} using the limits
\begin{equation}
s_c - R \cos \alpha_c \cos \phi_c \sin \delta \leq s \leq R
\label{A25}
\end{equation}
for $s$ and using eq. \ref{A24} to define the corresponding limits of $\theta$.

%\begin{figure}
%\vspace{7cm}
% \begin{center}{
%   \epsfxsize 5.0 true cm
%    \leavevmode
%   \epsffile{figa6.eps}
%  }\end{center}
%\caption{Coordinate system for partially visible spot}
%\end{figure}

\label{lastpage}


\begin{thebibliography}{10}

\bibitem{Albrow01a}
Albrow M., et al., 2001b, ApJ, 549, 759

\bibitem{Albrow01b}
Albrow M., et al., 2001a, ApJ, 550, 173

\bibitem{Alcock97}
Alcock C., et al., 1997, ApJ, 486, 697

\bibitem{Bogdanov95}
Bogdanov M.B., Cherepashchuk, 1995, Astron. Rep., 39, 873

\bibitem{Bogdanov96}
Bogdanov M.B., Cherepashchuk, 1996, Astron. Rep., 40, 713

\bibitem{Bontz79}
Bontz R.J., 1979, ApJ, 233, 402

\bibitem{Bryce00a}
Bryce H. M.,  Hendry M. A., Valls-Gabaud D., 2002, in 
'Microlensing 2000: A New Era of Microlensing Astrophysics', 
J.W. Menzies and P.D. Sackett eds., ASP Conf. Ser. 239, 195
(ASP: San Francisco) 

\bibitem{Dorren87}
Dorren J.D., 1987, ApJ, 320, 756

\bibitem{GauGou97a}
Gaudi B. S., Gould A., 1997, ApJ, 482, 83

\bibitem{GilDup96}
Gilliland R. L., Dupree A. K., 1996, ApJ, 463, L29

\bibitem{Gould94}
Gould A., 1994, ApJ, 421, 71

\bibitem{Gould952}
Gould A., 1995, ApJ, 441, L21

\bibitem{Gould97}
Gould A., 1997, ApJ, 483, 989

\bibitem{GouWel96}
Gould A., Welch D. L., 1996, ApJ, 464, 212 

\bibitem{Gray92}
Gray D.F., 1992, The Observation and Analysis of Stellar
Photospheres, (CUP: Cambridge)

\bibitem{Gray00}
Gray N., Coleman I. J., 2002, in 
'Microlensing 2000: A New Era of Microlensing Astrophysics', 
J.W. Menzies and P.D. Sackett eds., ASP Conf. Ser. 239, 204
(ASP: San Francisco) 

\bibitem{Han2000}
Han C., Park S.-H., Kim H.-I., Chang K., 2000, MNRAS, 316, 97

\bibitem{Hau99a}
Hauschildt P.H., Allard F., Baron E., 1999,
ApJ, 512, 377

\bibitem{Hau99b}
Hauschildt P.H., Allard F., Ferguson J., Baron E., Alexander D.R., 1999,
ApJ, 525, 871

\bibitem{Hendry98}
Hendry M. A., Coleman I. J., Gray N., Newsam, A. M., Simmons,
J. F. L., 1998, New Astronomy Reviews, 42, 125.

\bibitem{HeyLoe97}
Heyrovsk\'{y}, D., Loeb, A., 1997, ApJ, 490, 38

\bibitem{HS00}
Heyrovsk\'{y} D., Sasselov D., 2000, ApJ, 529, 69 

\bibitem{HeySasLoe99}
Heyrovsk\'{y} D., Sasselov D., Loeb A., 2000, ApJ, 543, 406 

\bibitem{Ignace99}
Ignace R. Hendry M. A., 1999, Astron. Astrophys., 341, 201

\bibitem{K94}
Kurucz R.L., 1994, ATLAS9 CDROMs

\bibitem{MaozGou94}
Maoz D.,  Gould A., 1994, ApJ, 425, 67

\bibitem{MenSac00}
Menzies J. W. and Sackett P. D., (eds.), 2001, 
Microlensing 2000: A New Era of Microlensing Astrophysics. ASP
  Conf. Ser. in press (ASP: San Francisco) 

\bibitem{NemWic94}
Nemiroff R. J., Wickramasinghe W., 1994, ApJ, 424, 21

\bibitem{Pac96}
Paczy\'{n}ski B., 1996, Ann. Rev. Aston. Astrophys., 34, 419

\bibitem{Peng97}
Peng E. W., 1997, ApJ, 475, 43

\bibitem{Sas97}
Sasselov D.,  1997, in  Variable Stars and the
Astrophysical Returns of the Microlensing Surveys, 
ed. R.~Ferlet et al., (Paris: Ed. Fronti\`eres), p. 141

\bibitem{Simmons95b}
Simmons J.F.L., Newsam A.M., Willis J.P., 1995, MNRAS, 276, 182

\bibitem{Simmons95a}
Simmons J. F. L., Willis J. P., Newsam A. M., 1995, A \& A, 293, L46

\bibitem{IAU176}
Strassmeier K.G., Linsky J.L. (eds), 1996, IAU Symp. 176, Stellar
surface structure, (Dordrecht: Kluwer)

\bibitem{Uitenb98}
Uitenbroek H., Dupree A. K., Gilliland R, L., 1998, ApJ, 116, 2501

\bibitem{VG94}
Valls-Gabaud D., 1994, in  Large Scale
Structures in the Universe, eds. J.~M\"{u}cket et~al., 
(Singapore: World Scientific), p. 326

\bibitem{VG96}
Valls-Gabaud D., 1996, in  Astrophysical applications of
gravitational lensing, IAU Symp 176, eds. C. S. Kochanek and
J.N. Hewitt, (Dordrecht: Kluwer), p. 237

\bibitem{VG98}
Valls-Gabaud D., 1998, MNRAS, 294, 747

\bibitem{Vermaak00}
Vermaak P., 2000, MNRAS, 319, 1011

\bibitem{WittMao94}
Witt H.A., Mao S., 1994, ApJ, 430, 505

\bibitem{Witt95}
Witt H.A., 1995, ApJ, 449, 42

\end{thebibliography}
\end{document}